\documentclass[%
 amsmath,amssymb,
reprint,%
superscriptaddress,
]{revtex4-2}

\usepackage{graphicx}
\usepackage{dcolumn}
\usepackage{bm}

\usepackage[utf8]{inputenc}
\usepackage[T1]{fontenc}
\usepackage{mathptmx}
\usepackage{etoolbox}

\usepackage{xcolor}

\newcommand{\degC}{{}^\circ\text{C}}
\newcommand{\PF}{{\cal P}}
\newcommand{\red}[1]{\textcolor{black}{#1}}
\newcommand{\rem}[1]{\iffalse {#1} \fi}

\makeatletter
\def\@email#1#2{%
 \endgroup
 \patchcmd{\titleblock@produce}
  {\frontmatter@RRAPformat}
  {\frontmatter@RRAPformat{\produce@RRAP{*#1\href{mailto:#2}{#2}}}\frontmatter@RRAPformat}
  {}{}
}%
\makeatother
\begin{document}

\preprint{AIP/123-QED}

\title[Exceptional Thermoelectric Power Factors in Hyperdoped, Fully Dehydrogenated Nanocrystalline Silicon Thin Films]{Exceptional Thermoelectric Power Factors in Hyperdoped, Fully Dehydrogenated Nanocrystalline Silicon Thin Films}
\author{Dario Narducci}%
\affiliation{\red{UdR INSTM of Milano Bicocca and} University of Milano Bicocca, Dept.\ Materials Science, via R. Cozzi 55, I-20125 Milan, Italy}
\email{dario.narducci@unimib.it.}
\author{Laura Zulian}%
\affiliation{University of Milano Bicocca, Dept.\ Materials Science, via R. Cozzi 55, I-20125 Milan, Italy}
\author{Bruno Lorenzi}%
\affiliation{University of Milano Bicocca, Dept.\ Materials Science, via R. Cozzi 55, I-20125 Milan, Italy}
\author{Federico Giulio}%
\affiliation{University of Milano Bicocca, Dept.\ Materials Science, via R. Cozzi 55, I-20125 Milan, Italy}
\author{Elia Villa}%
\affiliation{\red{UdR INSTM of Milano Bicocca and} University of Milano Bicocca, Dept.\ Materials Science, via R. Cozzi 55, I-20125 Milan, Italy}

						
\vspace*{-5cm}\hspace{-.8\columnwidth}\fbox{Cite as \textit{Appl. Phys. Lett.} \textbf{119}, 263903 (2021); https://doi.org/10.1063/5.0076547}

\begin{abstract}
Single-crystalline silicon is well known to be a poor thermoelectric material due to its high thermal conductivity. Most excellent research has focused on 
ways to decrease its thermal conductivity while retaining acceptably large power factors (PFs). Less effort has been spent to enhance the PF in poly- and nanocrystalline silicon, instead. Here we show that in boron-hyperdoped nanocrystalline thin films PF may be increased up to 33 mW K$^{-2}$m$^{-1}$ at 300 K when hydrogen embedded in the film during deposition is removed. The result makes nanocrystalline Si a realistic competitor of Bi$_2$Te$_3$ for low-temperature heat harvesting, also due to its greater geo-availability and lower cost.
\end{abstract}

\maketitle

Extended application of thermoelectric technologies either as heat harvesters or coolers critically depends on materials efficiency and geo-availability. Today, most thermoelectric devices make use of bismuth telluride and of its alloys, which grant a thermoelectric figure of merit of $\approx 1$ around room temperature. However, tellurium is well known to be scarce, with a geo-abundance comparable to that of platinum; and it is also a key raw material for competing technologies, such as photovoltaics and high-speed electronic devices. As a result, its price is discouragingly volatile, a second issue that makes large exploitation of thermoelectric technologies risky.
Aim of this \textit{Letter} is to show how nanocrystalline silicon may offer alternate paths to thermoelectric applications around room temperature.
\red{Although heat harvesting has mostly targeted high-enthalpy sources, it has been remarked \cite{Petsagkourakis2018,Haras2018} how microharvesting of low-temperature heat is of paramount importance to power sensing nodes (and low-power actuators) in the Internet of Things. Power outputs in the order of milliwatts are already at reach for commercial thermoelectric generators (TEGs), and the availability of low-cost TEGs based on silicon or other abundant raw materials might provide renewable powering solutions that could make IoT nodes fully maintenance-free \cite{Narducci2019}.   
}

Search for alternate thermoelectric materials has been the subject of an intense and largely diversified research. 
Among possible candidates, silicon has played since the Nineties a special role. While, upon suitable doping, it is possible to obtain values of electrical conductivity $\sigma$ and of Seebeck coefficient $\alpha$ leading to relatively high thermoelectric power factors $\PF=\sigma\alpha^2$, Si thermal conductivity $\kappa$ is unfortunately very large, resulting in meager thermoelectric figures of merit $zT=\PF T/\kappa$ (where $T$ is the temperature) of about 0.01 around room temperature. 
Despite its low efficiency, single-crystalline Si thin films have found niches of application in microelectronics because of their easy integration \red{\cite{Strasser2002,Yanagisawa2020}}. 

This state of affairs suddenly changed when thermoelectricity met nanotechnology. Dimensionally-constrained Si systems (nanolayers and nanowires) \red{enabled the manipulation of phonon transmissivity, leading to major reduction of $\kappa$}, from 140 W/mK down to the Casimir's limit \cite{Ju1999,nature1,nature2}. Then, the resulting $zT$ raised up to $\approx 0.8$ in single-crystalline Si nanowires. A major effort to convert such a proof-of-concept into viable technologies has arisen \cite{Li2011a,Fonseca2016,Tomita2018,Elyamny2020,Yanagisawa2020}, reporting significant results. Defect engineering \cite{Bennett2015} and holey silicon structures \cite{Ma2020} have also shown great opportunities in Si thin films.

As an alternate route, one may consider also the possibility of promoting Si efficiency moving from nanocrystalline samples, already showing low thermal conductivity because of phonon scattering at grain boundaries (GBs), and attempting to increase their power factors. Several strategies have been considered, including modulation doping \cite{Zebarjadi2011,Yu2012} and energy filtering. 
\red{
Energy filtering was found to occur in many materials, leading to non-standard simultaneous increase of $\sigma$ and $\alpha$ (Ref.\ \onlinecite{Narducci2015} and references therein). 
}

\red{
In previous publications we reported how, upon extended annealing at temperatures $\ge 800 \degC$, heavily boron-doped nanocrystalline silicon films could display an unexpected concurrent increase of their Seebeck coefficient and of their electrical conductivity \cite{Narducci2012}. We could correlate the increase of the power factor (PF) with the precipitation of 
boron (possibly as SiB$_3$) at GBs \cite{Narducci2014}. Precipitates set a double potential barrier that filters charge carriers, therefore enabling hot carriers only to diffuse upon application of a thermal gradient. As a result, mobile carriers move \emph{as if} no grain boundary were present while cold carriers are trapped within grains. Thus, mobile carrier density decreases, causing an increase of the Seebeck coefficient. At the same time their mobility increases since mobile carriers overcoming the potential barrier have energies larger (in absolute value) than thermalized holes. For the process to be beneficial, however, grain size must be small compared to carrier mean-free path, letting holes move in a semi-ballistic regime \cite{Narducci2012,Suriano2015}. 
The concurrent increase of Seebeck coefficient and electrical conductivity is witnessed by the temperature dependence of carrier mobility and density \cite{Zianni2019}. Mobility reports a single-crystal-like trend, since, as anticipated, mobile carriers overcoming the barriers are not scattered. Carrier density decreases in samples annealed above 800 ${^\circ}$C, consistently with the filtering mechanism \cite{Zulian2016}.
}
That such a model may explain the increase of the PF upon annealing was further corroborated by theoretical modeling and computational simulations \cite{Neophytou2013a,Neophytou2014,Zianni2015}. 

\begin{figure}
\includegraphics[width=\columnwidth]{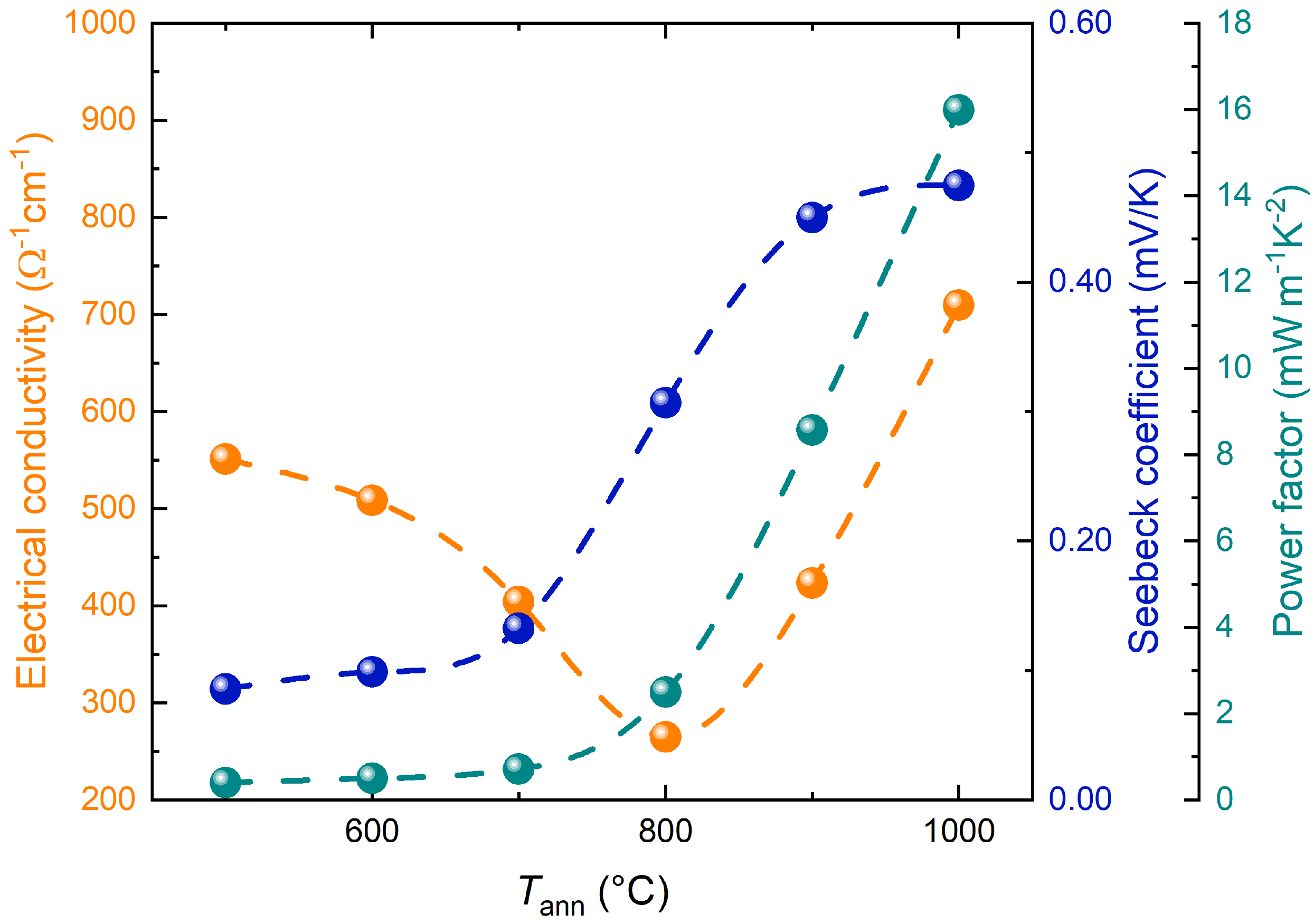}
\caption{Room temperature electrical conductivity, Seebeck coefficient, and power factor of nanocrystalline Si films upon subsequent two-hour annealing up to $T_\mathrm{a}$. Lines are for eye guide. Reproduced with adaptations from  Ref.\ \onlinecite{Narducci2015} with permission from the Royal Society of Chemistry.}
\label{fig:1}
\end{figure}

Figure \ref{fig:1} summarizes the trend of $\sigma$ and $\alpha$ observed upon sequential annealing in nanocrystalline silicon thin films, reporting a final power factor of 16 mW K$^{-2}$m$^{-1}$. 
\red{
For comparison, state-of-the-art Bi$_x$Sb$_{2-x}$Te$_3$ bulk alloys display PFs ranging between 3 and 5 mW K$^{-2}$m$^{-1}$ at 300 K (Refs.\ \onlinecite{Lee2010,Son2013,Kim2015}).
}

In this \textit{Letter} we discuss the role played by hydrogen in disabling energy filtering in hyperdoped nanocrystalline silicon films, further showing how full deydrogenation may lead to power factors of 33 mW K$^{-2}$m$^{-1}$. This may also provide an explanation of why, although heavily doped nanocrystalline silicon thin films are quite largely used in microelectronic industry for short-range electrical interconnects \cite{Kamins2012book}, no evidence of improved PFs had been reported in past years. 

Silicon thin films were grown onto 200-mm Si (100) wafers on which a 100-nm SiO$_x$ film was previously deposited. Nanocrystalline silicon films, 230-nm thick, were obtained by low-pressure chemical vapor deposition (LPCVD) at 625 $\degC$ using silane as reactive gas. A screen oxide (100 nm) was then deposited on top of the film and the wafers were submitted to blank B$^+$ ion implantation (26 keV, $6.0\times 10^{15}$ cm$^{-2}$ $+$ 47 keV, $6.0\times 10^{15}$ cm$^{-2}$) leading to a total nominal boron concentration of $4.4\times 10^{20}$ cm$^{-3}$. Boron was activated by Rapid Thermal Processing (RTP, 1050 $\degC$, 40 s). Finally, the screen oxide was removed by wet etch.
Annealing was performed in argon at a temperature $T_\mathrm{a}$ of  1000 $\degC$ for two hours on samples of different sizes. 
\emph{After annealing}, all samples were cut to give $50\times 5$ mm$^2$ chips and, following standard wet cleaning, Al pads were evaporated using shadow masks.  Seebeck voltages were measured using the integral method \cite{Wood} by fixing the temperature of the cold contact using a water--ice thermostatic bath while scanning the temperature of the other contact between 40 and 120 $^\circ$C. 
Hall resistivity was measured from 10 K to 300 K at a maximum magnetic field of 0.5 T on square $17\times 17$ mm$^2$ chips using evaporated Al contacts in a standard van der Pauw geometry. Resistivity was also determined by four-probe measurements after checking linearity of current-voltage characteristics.
Each set of measurements was repeated on the same sample at least three times to ensure data reliability and repeated on several nominally identical samples to verify materials reproducibility. Overall reproducibility was better than $\pm 6$\% on $\sigma$, $\pm 12$\% on $\alpha$, and $\pm 5$\% on Hall resistivity. Thus, PFs were obtained with a precision better than $\pm 30$\%. 
Both p and n-type single-crystalline silicon substrates of various resistivity were used to verify that SiO$_x$ punch-through had not affected measurements. Furthermore, substrates were grounded to prevent any polarization effect.

Fourier-Transform Infrared reflectance spectra were recorded using a Bruker Vertex 70v spectrometer in vacuum at room temperature with a resolution of 2.5 cm$^{-1}$ at an incidence angle of $45^\circ$. 

Transmission Electron Microscopy analyses\cite{Narducci2014} reported a columnar micromorphology  with grains extending on the whole film thickness, with in-plane (cross sectional) size of $\approx 50$ nm. Since boron pins grain 
boundaries\cite{boronsegregationinpolysilicon}, no change of grain size was observed upon annealing at $1000 \degC$ for 2 hours, in accordance with previous findings \cite{Narducci2014}.  

Comparison of $\sigma$ and $\alpha$ measured at 300 K before and after annealing reports a striking difference among samples depending on their aspect ratio (Table \ref{tab:size}). All as-deposited samples showed the expected low PF, as carrier scattering at GBs causes low electrical conductivity. Instead, after annealing, while samples of smaller sizes ($50\times 5$ mm$^2$, hereafter S-type) reported a large increase of their power factors, larger samples ($25\times 100$ mm$^2$, L-type) showed no significant change of $\sigma$ and $\alpha$. The same was observed when the whole wafer (200-mm diameter, W-type) was submitted to the heat treatment.

\begin{table*}[t]
\caption{Summary of samples processing conditions and of measured transport properties. 
S and L label specimens with sizes of $5\times 50$ 
and $25\times 100$ mm$^2$, respectively, while W specifies annealing carried out on a whole 200-mm wafer.  
Relative PFs are referred to the non-annealed and non-aged sample.}
\label{tab:size}
\begin{ruledtabular}
\begin{tabular}{rrllrrrrrr}
\multicolumn{3}{c}{Annealing conditions} & & & & & & \\
\cline{1-3}
Temperature                     & Duration                  & Size  & Aging & Conductivity                    & \multicolumn{1}{c}{Thermopower}            & \multicolumn{1}{c}{Hall mobility}                 & \multicolumn{1}{c}{Carrier Density}  & \multicolumn{1}{c}{PF} & \multicolumn{1}{c}{Relative}\\
					\multicolumn{1}{c}{($^\circ$C)} & \multicolumn{1}{c}{(min)} &       &       & ($\Omega^{-1}\mathrm{cm}^{-1}$) & \multicolumn{1}{c}{($\mathrm{mV K}^{-1}$)} & \multicolumn{1}{c}{(cm$^2$V$^{-1}$s$^{-1}$)}      & \multicolumn{1}{c}{(cm$^{-3}$)}          & (mW K$^{-2}$ m$^{-1})$ & PF\\
\hline
--   & --  & n/a & no & 562 & 0.16 & 13.5 &$2.60\times 10^{20}$ & 1.44 & 1\phantom{.00}\\
1000 & 120 & S   & no & 544 & 0.55 & 40.0 &$8.50\times 10^{19}$ & 16.5\phantom{0} & 11.4\phantom{5}\\ 
1000 & 120 & L   & no & 169 & 0.23 & 17.0 &$6.20\times 10^{19}$ & 0.89 & 0.62\\
1000 & 120 & W   & no & 223 & 0.21 &      &                     & 0.98 & 0.68\\
--   & --  & n/a & yes& 556 & 0.13& 14.0 &$2.30\times 10^{20}$ & 1.00 & 0.69\\
1000 & 120 & S   & yes& 588 & 0.75 & 57.0 &$6.30\times 10^{19}$ & 33.1\phantom{0}& 23.0\phantom{1}\\
1000 & 120 & L   & yes& 357 & 0.65 & 11.9 &$2.20\times 10^{20}$ & 15.1\phantom{9}& 10.5\phantom{0}\\
\end{tabular}
\end{ruledtabular}
\end{table*}

\begin{figure}
\includegraphics[width=\columnwidth]{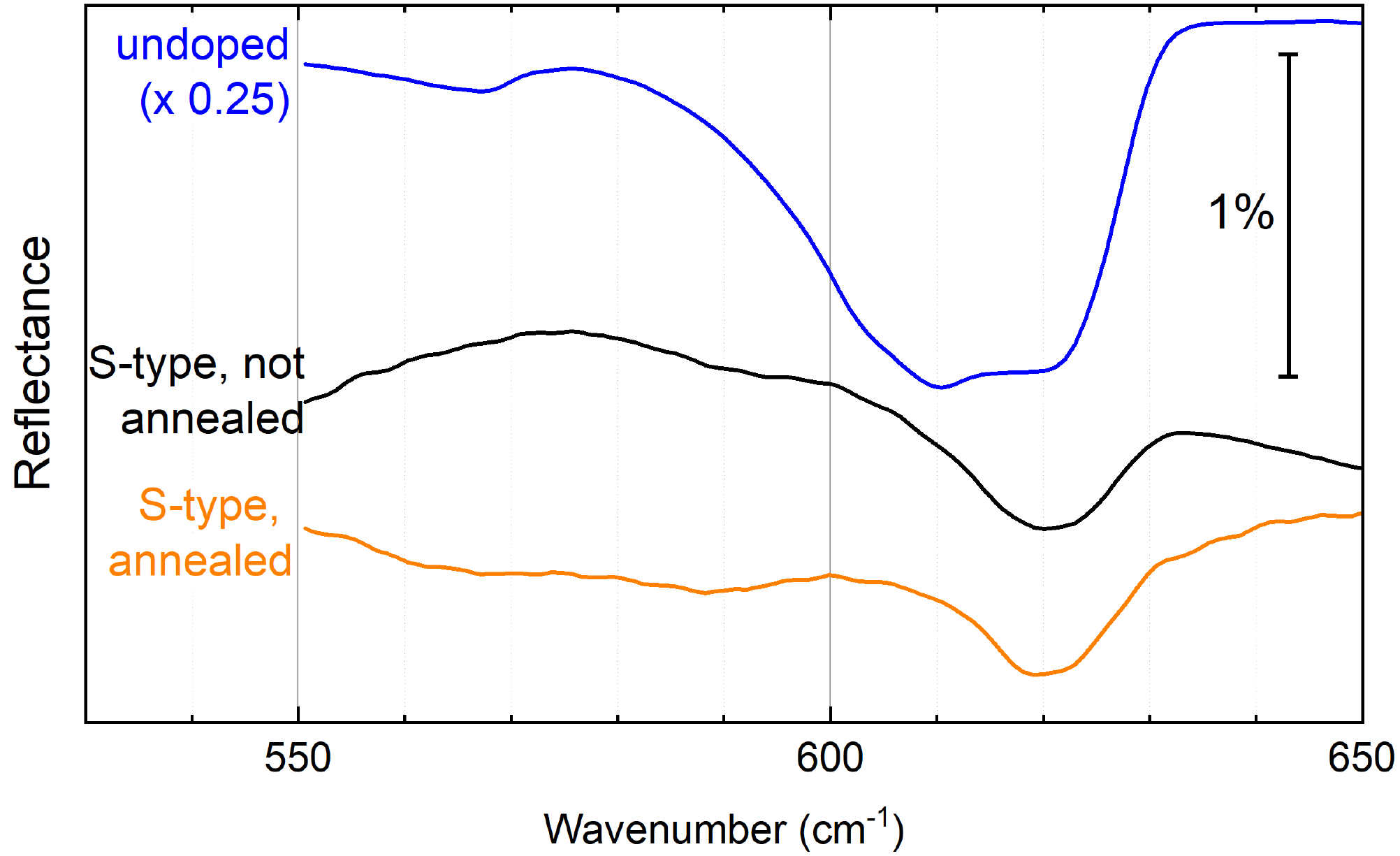}
\caption{Comparison among FTIR reflectance spectra of undoped and doped thin films before and after annealing  in the frequency region of Si--H wagging mode. \red{
In undoped Si, the largest hydrogen content, with H mostly bonded to Si, leads to a strong Si-H wagging signal. Upon boron implantation, a fraction of H leaves Si forming BH$_x$ complexes, weakening  the Si-H signal. Annealing (1000 $^\circ$C, 2 hours) causes hydrogen outdiffusion, both from BH$_x$ complexes and Si, further weakening the signal.  
} Spectra are displaced for clarity of presentation.}
\label{fig:IR}
\end{figure}

The role played by sample shape at the millimetric scale is rather puzzling, yet not unique. Solovyov and coworkers reported a similar dependence of materials characteristics when forming 
superconducting YBa$_2$Cu$_3$O$_7$ (YBCO) from pre-deposited Y, Cu, and BaF$_2$ films \cite{Solovyov2001}. The process required precursor oxidations in a O$_2$/H$_2$O stream, leading to the release of HF. It was shown that the efficiency of HF removal critically depends on the sample size because, in a laminar flow, a stagnating HF pressure may establish on long specimens, preventing the conversion of the pristine film into YBCO. A comparable mechanism may occur in the present system. 
Hydrogen is known to be embedded in silicon upon CVD from SiH$_4$. Its residual concentration largely depends on the deposition temperature, being greater for lower substrate temperatures. However, also around 620 $\degC$ its presence in polycrystalline silicon has been reported \cite{Kamins1982a,Kamins1982b}. 
The presence of hydrogen in our films was confirmed by infrared reflectance spectroscopy. Figure \ref{fig:IR} compares the spectra of non-implanted (undoped) and implanted (boron-doped) nanocrystalline Si films before and after annealing.   
Signature of wagging modes of vibrations of the mono-hydride (Si--H) \cite{Bakr2011,Waman2011,Jadhavar2017} is observed at $\approx 620$ cm$^{-1}$. 
Note that its intensity decreases in the sequence, being larger in the undoped film, smaller after doping and further decreasing after annealing. 
\red{
In undoped Si, hydrogen is largely bonded to Si, leading to the strongest Si-H wagging signal. When boron is implanted (and the film is subjected to RTP only), a fraction of H leaves Si and preferentially binds to boron, forming BH$_x$ complexes. Thus, Si-H signal weakens. Annealing at 1000 $^\circ$C, 2 hours in S-type samples leads to hydrogen outdiffusion, both from BH$_x$ complexes and Si (\textit{vide infra}), further weakening Si-H wagging signal.  
}
Signals in the 2000-2150 cm$^{-1}$ range (not shown) could not be unambiguously assigned, since in that frequency region interference fringes due to multiple reflection at the film--oxide interface overlap \cite{Herrero1991}.

Removal of hydrogen by annealing occurs while flowing Ar. 
Upon heat treatments under 
an Ar flux of $4.26\times 10^{-6}$ m$^3$/s (at room temperature), 
gas velocity $v$ at 1000 $^\circ$C computes to $1.44\times 10^{-2}$ m/s so that the Reynolds number is found to be 3.35 -- largely within the laminar regime. Critical entry length is then 0.46 mm, displaying that the flux is fully developed well before reaching the samples, placed about 30 cm from gas entry. 
Gas flow is effective at promoting the outdiffusion of H$_2$ from the sample when it prevents the formation of a stagnating H$_2$ layer in proximity of the film surface. Hydrogen flux desorbing from the sample surface follows Fick's equation that, in the presence of a source, reads \cite{Solovyov2001}
\begin{equation}
D_\mathrm{H_2}\nabla^2 p_\mathrm{H_2} -v\nabla p_\mathrm{H_2}=0
\end{equation}
where $D_\mathrm{H_2}$ is the hydrogen diffusivity in the gas phase at the operative pressure ($\approx 10^5$ Pa) and temperature.  For low $v$, the second term in the previous equation (often referred to as the 'convection term') may be neglected \cite{Solovyov2001}. Thus, hydrogen flux leaving the sample computes to \cite{Holman1986}
\begin{equation}
\Phi_\mathrm{out}=D_\mathrm{H_2} S_\mathrm{f}\Delta p_\mathrm{H_2}/k_\mathrm{b}T
\end{equation}
where $k_{\rm b}$ is the Boltzmann constant, $\Delta p_\mathrm{H_2}$ is the difference between H$_2$ partial pressures at sample surface $p^\mathrm{(s)}_\mathrm{H_2}$ and in the chamber $p^\mathrm{(a)}_\mathrm{H_2}$ while $S_\mathrm{f}$ is a shape factor,
\begin{equation}
S_\mathrm{f}=\frac{2\pi L}{\ln(4L/W)}   
\end{equation}
where $L$ and $W$ are the sample length (parallel to the chamber axis) and width. 
For the sample sizes considered in this work one computes $S_\mathrm{f}$ to be 8.52  
and 22.6 cm for S  
and L type, respectively. If one sensibly assumes that the Ar flux $\Phi_\mathrm{Ar}$ must exceed $\Phi_\mathrm{out}$ to effectively strip away  hydrogen, then Ar flux that may prevent a stagnating H$_2$ layer at S-type sample surface may not be sufficient to that aim in L-type (and W-type) specimens.
Therefore, $\Phi_\mathrm{Ar}$ may be insufficient to prevent the formation of a stagnating H$_2$ layer in proximity of the specimen surface when its length increases. 

To check this hypothesis, L-type samples were submitted to annealing for two hours at 1000 $^\circ$C under a tripled Ar flux. Under such conditions we found that S- and L-type PFs aligned to each other, suppressing the size effect. L-type specimens reached a PF at 300 K of $23.4 \pm 7.0$ mW K$^{-2}$m$^{-1}$ 
while S-type samples annealed under the same conditions reported a PF of $18.5 \pm 5.5$ mW K$^{-2}$m$^{-1}$. 

The presence of hydrogen in the films along with its critical role to enable energy filtering suggests that hydrogen interferes with boron precipitation. 
Hydrogen incomplete removal from the film prevents the formation of SiB$_x$ precipitates that are responsible for the formation of the potential barriers needed to enable energy filtering. 
Thus, during the extended annealing, two processes are stipulated to occur. Hydrogen diffuses to the film surface, as its concentration exceeds its solubility in Si at any temperature \cite{Rath2000} -- but it is stripped away by the Ar stream only when no hydrogen stagnating layer forms at film surface. When hydrogen may be factually removed, 
boron in excess to its solubility concentration diffuses to reach GBs, where it preferentially precipitates as SiB$_x$ 
and the potential barriers responsible for the enhancement of the PF may establish.

Three pieces of evidence are advanced to support this model. 
First, one  observes that 
precipitation at GBs may occur only at temperatures such that boron diffusion length $L_\mathrm{D}=2\sqrt{D_\mathrm{B}(T_\mathrm{a})\tau_\mathrm{a}}$ (where $\tau_\mathrm{a}$ is the duration of the annealing) is larger than the grain size $\ell$. Boron diffusivity depends on temperature\cite{Horiuchi1975} as
$D_\mathrm{B}(T)=(6.01\times 10^{-3} \mathrm{cm^2/s}) \exp(-2.51 \mathrm{eV}/k_\mathrm{b}T)$, 
so that for $\tau_\mathrm{a}= 7.2\times 10^3$ s the lowest annealing temperature enabling migration of boron to grain boundaries is $\approx 712 \degC$. 
This is consistent with the fact that annealing at lower temperatures never leads to a change of the PF, irrespective of the sample size.

If hydrogen is the limiting factor to boron precipitation, than 
any process enabling a fuller sample dehydrogenation might lead to additional boron precipitation and to larger PFs. 
That hydrogen is not fully removed by annealing, even in S-type samples, is witnessed by FTIR spectra (Fig.\ \ref{fig:IR}).
Longer annealing times ($\gg 2$ hours) were then attempted to completely deplete the films. However, extended thermal treatments led to an increase of grain size. Since energy filtering requires grain sizes smaller than the carrier mean-free path, grain growth makes the system leave the semi-ballistic regime.
However, as mentioned, hydrogen supersaturates silicon at any temperature. Thus,  
sample aging may show effective at enabling dehydrogenation since, even under static conditions (no flux), the surface H$_2$ layer has the time to diffuse out. If this is the case, dehydrogenation should occur independently of the specimen size.

Samples aged at room temperature for about five years displayed unchanged PF before annealing. Instead, after annealing at $1000 \degC$ (as required to let boron diffuse to the grain boundaries), they reported (Table \ref{tab:size}) an enhanced PF. Both S- and L-type samples displayed improved power factors, which were larger in S-type samples, where \red{the value} of 33 mW K$^{-2}$m$^{-1}$ was attained -- almost twice the record PF of YbAl$_3$ \cite{Lehr2013}. In terms of power output this implies that, 
\red{upon application of a (modest) in-plane temperature difference of 10 K along a $1\times 1$ cm$^2$ planar TEG of thickness $d = 200$ nm, an output power density $\PF(\Delta T)^2/4d$ of 8.25 mW/cm$^2$ is predicted.} 
In-plane thermal conductivity measurements are under way. However, one may \emph{estimate} that for in-plane grain size of 50 nm, thermal conductivity should be of order 10 W/mK (Ref.\ \onlinecite{McConnell2005}). \red{Although no claim of $zT$ can be made based on $\kappa$ literature data, were film thermal conductivity in the order of 10 W/mK, the material would show a figure of merit of $\approx 0.99$ at 300 K.} 

The detailed mechanism of hydrogen interference with boron precipitation remains to be elucidated. 
One may speculate that, since hydrogen forms complexes with boron \cite{Seager1991,Pearton2013}, it might subtract free boron from the precipitation equilibrium, therefore preventing or partially preventing its diffusion-limited precipitation \cite{Ham}. 
This might find a very preliminary support from the observed trend of the wagging mode intensity (Fig.\ \ref{fig:IR}), suggesting that upon doping (boron-implantation), mono-hydrides partially decompose to form complexes with boron, and then the total hydrogen content is further depleted by outdiffusion occurring during the annealing.
As an alternative, hydrogen might compete with boron at decorating GBs, therefore hindering its localized precipitation. At this stage, both mechanisms remains plausible, along with their simultaneous occurrence in the films. Further investigations are needed to provide clear evidence of the active mechanism(s).

\begin{figure}
\includegraphics[width=\columnwidth]{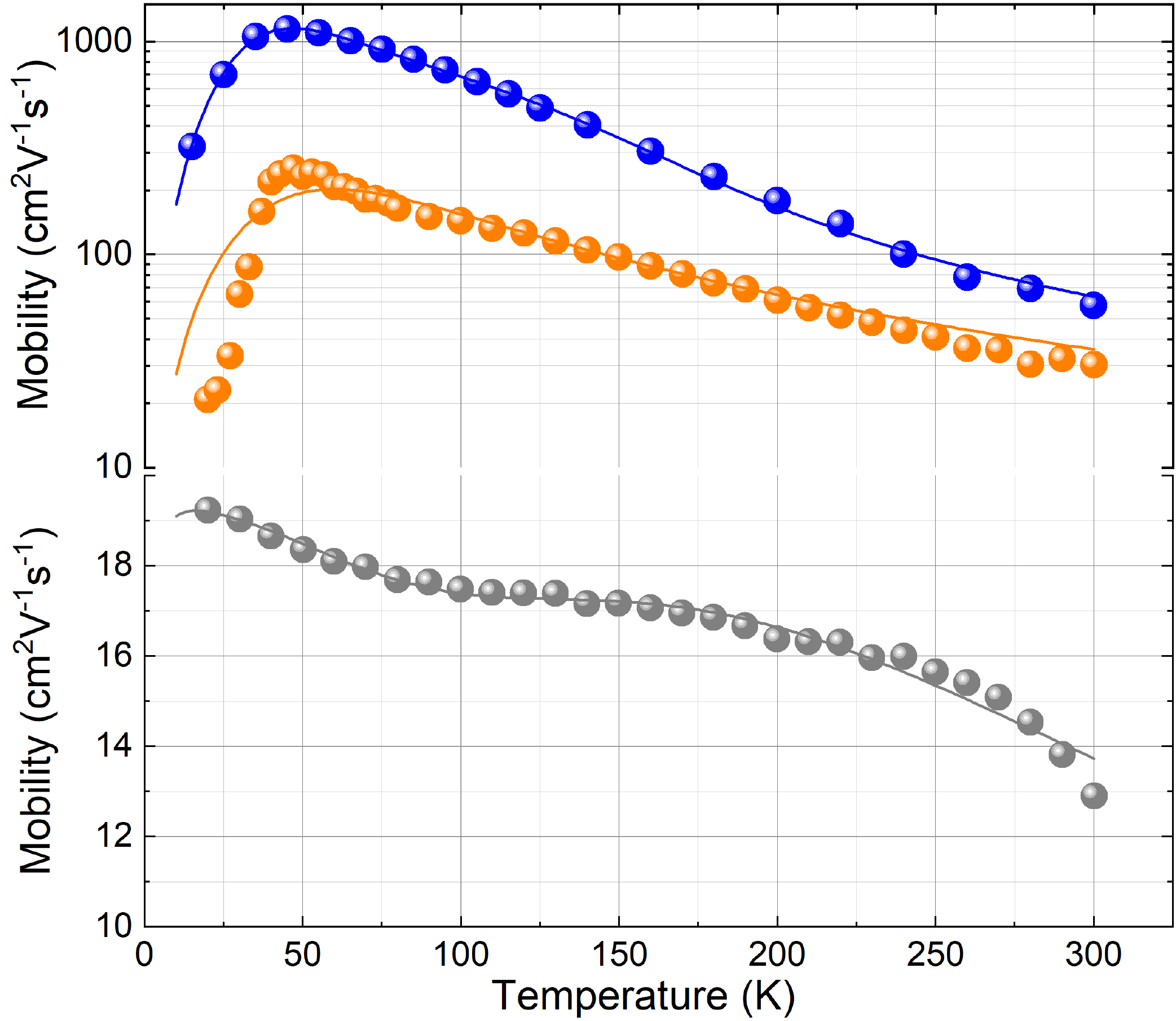}
\caption{Hall mobility vs.\ temperature of S-type samples, not annealed (gray) and annealed (1000 $\degC$, 2 hours) before (orange) and after (blue) aging, fitted to the pertinent scattering model.}
\label{fig:Hall}
\end{figure}

To confirm that the additional improvement of PFs originates from energy filtering, we examined Hall mobility $\mu$ as a function of temperature for S-type samples, untreated, and annealed at 1000 $\degC$, either before or after aging. 
(Fig.\ \ref{fig:Hall}).  
In general terms, mobility is limited by ionized impurity, electron-phonon, and GB scattering. Following Brooks-Herring's (BH) model \cite{BHModel}, when scattering by ionized impurities dominates then $\mu_\mathrm{BH}\propto T^{3/2}$. Electron-phonon scattering must account instead for both intravalley and intervalley scattering, with $\mu_\mathrm{e-ph}\propto T^{-2.1}$ in silicon \cite{LundstromBook}. Scattering by grain boundaries is instead more critical to model. When potential barriers $V_\mathrm{GB}$ are large, with boundaries inelastically scattering most carriers, Seto's model \cite{Seto} provides an adequate approximation, with $\mu_\mathrm{GB}\propto T^{-1/2}\exp(-eV_\mathrm{GB}/k_\mathrm{b}T)$ (where $-e$ is the electron charge). Matthiessen's rule let combine the three mechanisms. Fitting to experimental data shows that mobility in untreated samples is almost constant, with values around 15 cm$^2$V$^{-1}$s$^{-1}$ at 300 K, typical of materials wherein carrier mobility is limited by scattering at GBs. Instead, $\mu(T)$ of annealed samples is fully comparable to that of single crystals, being dominated by intravalley and intervalley electron-phonon scattering at high temperatures and by ionized impurity scattering at low temperatures \emph{as if} GBs were transparent to carriers. 
Thus, mobility is in accordance with the energy-filtering model. Low-energy (`cold') carriers are trapped between barrier pairs, being localized within the grain, not contributing to charge transport processes and therefore not contributing to $\mu$ either. Instead,  high-energy (`hot') carriers move above the barriers, seeing the medium as single-crystalline.

\red{
Since silicon thin films were the subject of several previous investigations, a comparison with results reported by other research groups over recent years may be useful, also to discuss why a similar major enhancement of the power factor could not be observed.
\\
It should be noted that extended annealing at high temperatures are not customary in silicon, so that only few investigations explored this avenue to increase Si PFs. 
Among those addressing post-deposition thermal treatments, 
several have focused on nanocomposites of Si nanocrystals embedded in a hydrogenated amorphous silicon (aSi:H) tissue. In a series of papers, Loureiro and co-workers \cite{Loureiro2015,Loureiro2017} discussed the effect of radio-frequency power density used to grow in-situ boron-doped films, reporting about an anomalous increase of the electrical conductivity at constant Seebeck coefficient, which was tentatively related to the occurrence of energy filtering due to boron precipitation at grain boundaries. However, due to the prevailing carrier scattering in the amorphous matrix, PF could not exceed 0.4 mW K$^{-2}$m$^{-1}$. Comparable results were obtained by Acosta \textit{et al.}\cite{Acosta2019}, who analyzed in detail the role played by hydrogen in their system. Also in this case, however, the best PF was limited by the presence of the amorphous tissue that, while positively impacting $zT$, limited the PF to $\approx 0.2$ mW K$^{-2}$m$^{-1}$. 
The effect of annealing on heavily boron-doped Si nanocrystals embedded in aSi:H  was finally reported by Zhang and co-workers
 \cite{Pham2020}.  Annealing at 600 $^\circ$C for 10 minutes of samples with a nominal B density $\ge 10^{20}$ cm$^{-3}$ led to the full recrystallization of the thin film and to the formation of grains with sizes smaller than 5 nm. In such system, violations of the standard Pisarenko relation was observed and tentatively explained as a result of energy filtering. Rather unfortunately, no annealing above 600 $^\circ$C and/or for longer times was attempted presuming that this would have led to grain growth. Therefore, the largest PF was of $\approx 2$ mW K$^{-2}$m$^{-1}$, yet a remarkably large value in nanocrystalline films, considering that neither full precipitation of boron or hydrogen outdiffusion could take place. 
\\
SiGe alloys were also considered. Although their electronic structure and physical chemistry obviously differ from that of silicon, a SiGe system quite similar to the one we have investigated was discussed by Li \textit{et al.} \cite{Lu2015}. Thin films of Si$_{74}$Ge$_{26}$ alloys were grown by LPCVD on a Si (100) wafer that was previously coated with a SiO$_x$ film and an amorphous Si thin layer. Upon ion implantation followed by rapid thermal annealing, the film was submitted to additional annealing at 1000 $^\circ$C for one hour. Film developed columnar grains with an average in-plane grain size of $\approx 275$ nm, largely exceeding the carrier mean free path. Subsequent carrier thermalization within grains may explain why in no case the films displayed PF improvements.
In a more recent study mostly focused on ternary Si$_{1-x-y}$Ge$_x$Sn$_y$ alloys \cite{Peng2019,Lai2021}, Si$_{1-x}$Ge$_x$ control samples doped with boron 
were also prepared and analyzed. While in this case grain size was suitable to enable energy filtering ($\approx$ 10 nm), rapid thermal annealing (15 s at temperatures up to 1150 $^\circ$C) was too short to promote boron precipitation at grain boundaries. Thus, 
only a very small improvement of the PF was reported (up to 0.1 mW K$^{-2}$m$^{-1}$), as no energy filtering apparently occurred.
\\
Comparison to the literature seemingly confirms that to observe energy filtering in heavily boron-doped polycrystalline Si the material must meet three requisites: (\textit{a}) dopant concentrations must exceed the solubility threshold at the annealing temperature; (\textit{b}) annealing conditions must be such that dopant diffusion lengths are at least comparable to the grain size, allowing excess boron to diffuse to grain boundaries; and (\textit{c}) grain size must be smaller than carrier mean-free path to prevent carrier thermalization. Should any of such conditions not met, no energy filtering may occur.
}

In summary, we have shown that hydrogen largely rules the possibility of enabling energy filtering in nanocrystalline silicon thin films heavily doped with boron. Only when hydrogen is effectively removed, annealing at 1000 $\degC$ induces boron precipitation at grain boundaries, enabling hole energy filtering. To this aim, proper processing conditions are needed, preventing the formation of stagnating H$_2$ layers at film surface. Aging additionally promotes hydrogen outdiffusion at room temperature. As a result, power factor may be largely increased, up to 33 mW K$^{-2}$m$^{-1}$ at 300 K -- an exceptionally large value for films where an in-plane thermal conductivity at room temperature of $\approx 10$ W K$^{-1}$m$^{-1}$ is anticipated.
Since hydrogen is an elusive impurity in silicon, its critical role in CVD-grown thin films may explain why, despite the remarkable research effort devoted to nanocrystalline Si films for thermoelectric applications, observations of enhanced PFs may have been missed in the past. 

\begin{acknowledgments}
\red{
Authors acknowledge the partial financial support received by the Ministry of the Economic Development (MiSE) through the collaboration contract "Exploitation of new active materials for the development of thermoelectric microgenerators", Project 1.3 "Frontier materials for energy uses" (CUP I34I19005780001), Three-year Implementation Plan 2019-2021. 
}
\end{acknowledgments}

\section*{Author Declarations}
\subsection*{Conflict of Interest}

The authors have no conflicts to disclose.

\section*{Data Availability Statement}

The data that support the findings of this study are available from the corresponding author upon reasonable request.

%
%
%

%

\end{document}